\documentclass [12pt] {article} 
\usepackage{epsf,youngtab} 
\begin{document}
\begin{titlepage}
\begin{flushleft}        
       \hfill                      USITP-00-01\\
      \hfill                       \\   	                                 
\end{flushleft}
\vspace*{3mm}
\begin{center}
{\LARGE Free Large \( N \) Supersymmetric \\ Yang-Mills Theory as a String 
Theory\\}
\vspace*{12mm}
{\large Parviz Haggi-Mani\footnote{E-mail: parviz@physto.se} and Bo Sundborg\footnote{E-mail: bo@physto.se} \\
        {\em Institute of Theoretical Physics \\
        Box 6730\\
        S-113 85 Stockholm\\
        Sweden\/}\\}
\vspace*{25mm}
\end{center}

\begin{abstract} 
The strong version of Maldacena's AdS/CFT
conjecture implies that the large \( N \) expansion of free \( {\cal N}=4 
\) super-YM 
theory describes an interacting 
string theory in the extreme limit of high spacetime curvature relative 
to the string length. 
String states may then be understood as composed of SYM 
string bits. We investigate part of the low-lying spectrum of the 
tensionless (zero-coupling) limit and find a 
large number of states that are not present in the 
infinite tension (strong-coupling) limit, notably several massless spin 
two particles. We observe that all conformal dimensions
are \( N \)-independent in the free SYM theory, implying that masses 
in the corresponding string theory are unchanged by string interactions. 
Degenerate string states do however mix in the 
interacting string theory because of the complicated \( N 
\)-dependence of general CFT two-point functions. Finally we verify 
the CFT crossing symmetry, which corresponds to the dual properties of 
string scattering amplitudes. This means that the SYM operator 
correlation functions define AdS dual models analogous to the Minkowski 
dual models that gave 
rise to string theory.
\end{abstract} 

\end{titlepage}

\section{Introduction}

Since 't Hooft's original discussion \cite{'tHooft:1974jz} of the large \( N \) behaviour 
of gauge theories we have had a picture of a topological expansion of 
gauge theories in terms of surfaces of different genus, resembling 
the genus expansion of string amplitudes. In recent years Maldacena's 
conjecture \cite{Maldacena:1997re}, relating the large \( N \) expansion of \( {\cal 
N}=4 \) supersymmetric Yang-Mills theory to string theory in an \( 
AdS_{5} \times S^5 \) background has stimulated a resurge of interest 
in the large \( N \) limit. The conjectured correspondence permits 
the calculation of previously inaccessible gauge theory quantities by means 
of classical supergravity techniques when the 't Hooft coupling, \( \lambda = 
g_{YM}^2 N \), is large 
\cite{Gubser:1998bc,Witten:1998qj}\footnote{For further references to 
these developments see the comprehensive review \cite{Aharony:1999ti}.}. Although this 
regime is a concrete 
realization of the Yang-Mills/String duality, the string theory side 
is somewhat crippled: only the lowest, massless string states 
contribute, and with few exceptions only tree level interactions have 
been investigated.

To study the relation between \emph{quantized strings} and gauge 
theory in the AdS/CFT setting, one has to consider intermediate or small \( \lambda 
\), and the limit of vanishing \( \lambda \) naturally presents itself 
as a manageable alternative zeroth order approximation. Then the string tension \( T  \sim 
\sqrt{\lambda}/R^2 \) effectively goes to zero, if the radius of 
curvature \( R \) of the background is kept fixed. Or the radius of curvature
becomes much smaller than the string 
length scale \( l_{s} \sim T^{-1/2} \), i.e. \( R/l_{s} \ll 1  \). 
There are arguments \cite{Kallosh:1998qs} to all
orders in the 
string coupling \( 
g_{s} = g_{YM}^2 \) and \( \alpha' = l_{s}^2 \) that the \( AdS_{5} 
\times S^5 \) background is a solution to string theory, and it seems 
natural to assume that \( \lambda = 0\) gauge theory is dual to (or 
can serve as a definition of) zero tension string theory on this 
background. Certainly, the two theories should both be symmetric under 
\( SU(2,2|4) \), acting as a superconformal group on the gauge theory, 
and as anti-deSitter supersymmetries on the string theory. Here we note 
that the problem of defining 
quantized tensionless or null strings in flat backgrounds 
\cite{null} is in fact a more complicated problem than the present 
AdS case, due to its lack of a curvature scale, and its 
solution relies on additional assumptions. 

Because the full \( AdS_{5} \times S^5 \) background also contains a 
Ramond-Ramond field which prohibits the use of conventional string quantization 
methods, the quantization is a very difficult problem. Although 
interesting progress has been made 
\cite{Pesando:1998fv,Kallosh:1998nx,Kallosh:1998ji,Berkovits:2000fe}, we 
propose a different 
route. We assume that the strong version of Maldacena's conjecture 
works, i.e. that the gauge theory describes string theory even at small 't 
Hooft coupling. Then we can ask whether the picture of string theory 
that emerges is consistent with general expectations about the 
behaviour of string theory. Thus, one can get indirect evidence for 
or against the strong form of the AdS/CFT correspondence, by 
collecting knowledge about the gauge theory, which can be 
interpreted as knowledge about string theory until evidence is found 
to the contrary. Since we have not found any such negative evidence we 
will use gauge theory and string theory terminology interchangeably, 
but it should be remembered that all our calculations are done in 
gauge theory.

String theory can usually be characterized by its asymptotic states 
and interactions between them encoded in the scattering matrix. In an 
AdS background one immediately runs into conceptual problems, since 
neither the notion of asymptotic states nor of an ordinary 
\( S \)-matrix are well 
defined. Still, in terms of perturbations on 
the boundary of AdS, Balasubramanian et. al. \cite{Balasubramanian:1999ri} and 
Giddings \cite{Giddings:1999qu} 
have argued for a kind of generalized \( S \)-matrix, which replaces 
the usual \( S \)-matrix for string theory in this background. It is also 
directly related to CFT correlation functions by the AdS/CFT correspondence. 

While we cannot isolate ordinary asymptotic states in an AdS background, 
we can do equally well, at least in principle. The spectrum (of energy in 
global coordinates) is discrete, and we could study how interactions 
affect the states of the theory. In the zero \( \lambda \) limit we are 
considering, this is a purely combinatorial problem. The leading three-point 
functions of the gauge-invariant states which admit a string interpretation 
are of order \( 1/N \sim \kappa/R^4\), where \( \kappa \) is the 
gravitational coupling. To leading order in large \( N \) 
single-string states can be viewed as covariant strings of 
super-Yang-Mills string 
bits\footnote{String bits have been proposed by Thorn \cite{Thorn:1991fv} as 
possible constituents of strings in a non-covariant formulation.}. 
These AdS 
states correspond to CFT states, and by the CFT operator-state correspondence 
we could find associated operators, which are the operators 
involved in the generalized \( S \)-matrix. 

For each AdS state there is 
a deformation of the string theory background 
\cite{Gubser:1998bc,Witten:1998qj,Aharony:1999ti}. The most important 
deformations are the relevant and marginal deformations, which do not ruin 
the UV properties of the CFT, or the asymptotically locally AdS nature of 
the corresponding spacetime. In section \ref{Sp} we list all such (primary)
operators composed exclusively of scalars. Surprisingly, we find several 
operators corresponding to massless spin two fields in the bulk. We also 
discuss how string states mix by \( 1/N \) corrections, and how the string 
propagator can be diagonalized. 

In string theory, all the essential information about interactions is 
encoded in the three-string vertex. Similarly, the interactions in the 
conformal field theory are summarized in the operator product 
expansion. Not surprisingly three-string vertices and the OPE correspond 
closely to one another in the AdS/CFT dictionary. In section \ref{Opsv} we study 
general features like selection rules in the \( \lambda = 0 \) case, to 
leading order in 
large \( N \), and also discuss some 
important special cases. We also dispel the fear that free field 
theory is too trivial to describe a complicated interacting string 
theory. 

Given a generalized S-matrix we may discuss the properties of amplitudes. 
Relativistic amplitudes should obey crossing symmetry, whether they 
are point-particle amplitudes or string amplitudes, but whereas
point-particle amplitudes can be obtained from sums of different Feynman 
diagrams with singularities in distinct crossed channels, string 
amplitudes come from string diagrams which by analytic continuation 
each exhibit singularities in several crossed channels. This property 
of string amplitudes was called ``duality'' in the early days of 
string theory. In section \ref{Cd} we check the crossing symmetry of a 
particular CFT four-point function, which translates to 
duality of the generalized string four-point amplitude. We also indicate 
a simple direct argument for general crossing symmetry in the kind of CFT 
built on free field theory that we are considering.

\section{States and propagators}
\label{Sp}

In addition to the gauge potential the \( {\cal N}=4 \) 
supersymmetric Yang-Mills theory contains six scalars in the adjoint 
representation of the gauge group, as well as fermions. Local conformal 
operators may be written as products of fundamental fields in the 
adjoint representation (the field strength in the case of the gauge 
potential). Covariant derivatives (ordinary derivatives at \( \lambda = 0 \))
on the fundamental fields are also 
allowed. When 
the trace of the product is taken one gets 
invariants. Because of the cyclic symmetry of the trace, we may think 
of the single-trace operators as necklaces (closed strings) composed 
of SYM beads (string bits). Multiple-trace operators, i.e. products 
of single-trace operators, correspond to multi-string states. In
\cite{Sundborg:1999ue} the full spectrum of single-trace fields in 
the zero \( \lambda \) limit is given, but in 
this paper we instead focus on some general features of 
correlation functions/string amplitudes. At zero
\( \lambda \) different fundamental fields propagate independently so it 
is perfectly 
consistent to 
restrict attention to a subset of them. For simplicity we only consider 
conformal operators built of the six scalar fields \( \phi^{I} \):
\begin{eqnarray}
	    (\partial^{{\{n}\}}\Phi^{\{I\}})_{\{\mu\}}
	        &\equiv &(\partial^{n_{1}\ldots n_{k}}\Phi^{I_1...I_k})
	            _{\mu^1_{1}\ldots\mu^{n_{1}}_{1}\ldots{\mu^1_{k}\ldots\mu^{n_{k}}_{k}}}   
	    \label{eq:} \nonumber \\
	     & \equiv & {1 \over N^{k/2}}
	        {\mathrm{Tr}} 
	        \left\{(\partial_{\mu^1_{1}}\ldots\partial_{\mu^{n_{1}}_{1}}\phi^{I_1}) 
	           \ldots(\partial_{\mu^1_{k}}\ldots\partial_{\mu^{n_{k}}_{k}}\phi^{I_k})\right\},
	        \label{eq:operators}
\end{eqnarray}
where we have introduced multiple indices denoted with braces. Note that 
Hermitean operators generally are special linear combinations of such 
operators.

We study operators of definite conformal dimension. In our simple 
setting without interactions, the dimension is additive. The 
fundamental scalar has dimension \( \Delta_{\phi}=1 \) and the 
derivative (the translation generator) has \( \Delta_{\partial}= 1 
\). Primary operators are operators which (at the origin) are annihilated by 
special conformal transformations. From them descendant operators, said 
to belong to the same conformal family, are 
created by repeated application of the other conformal generators, in 
effect the derivative. In the AdS picture 
the primary operator gives a ground state for the Hamiltonian conjugate 
to the global time coordinate, and the descendants are excited states, 
which may be obtained by acting with AdS isometries not commuting with 
the Hamiltonian. Thus all the particles in AdS can be listed by only 
listing the corresponding conformal primaries. It is also enough to consider 
the correlation functions of the
primaries, since those of descendants are related by
conformal symmetry.

The propagator of a scalar field in the adjoint representation of \( 
SU(N) \) is
\begin{equation}	
\langle\phi^{\alpha}_{\beta}(x)\phi^{\gamma}_{\delta}(y)\rangle=
(\delta^{\alpha}_{\delta}\delta^{\gamma}_{\beta}-\frac{1}{N}
\delta^{\alpha}_{\beta}\delta^{\gamma}_{\delta})|x-y|^{-2\Delta_\phi}
	\label{ScalarProp}
\end{equation}
where the first term is the only one for the group \( U(N) \), allowing 
for 't Hooft's double line representation \cite{'tHooft:1974jz} in the large \( N \) 
limit. For \( SU(N) \) the second term can be dealt with by \( 1/N \) 
corrections to the naive double line diagrams. The above propagator for 
fundamental scalars can be used to calculate any correlation function
\begin{equation}
    \langle \:\partial^{\{n_{1}\}}\Phi^{\{I_{1}\}}(x_{1})\;\:
    \partial^{\{n_{2}\}}\Phi^{\{I_{2}\}}(x_{2})\ldots\:
    \partial^{\{n_{m}\}}\Phi^{\{I_{m}\}}(x_{m})\:\rangle
	\label{}
\end{equation}
in the \( \lambda = 0 \) limit, e.g. by making all possible 
contractions directly, or by using Wick's theorem (all 
conformal operators are defined to be normal ordered). In particular 
any scalar two point function may be calculated, and the results give a metric 
in the space of operators, 
\begin{equation}
	\langle A(x_{i}) B(x_{j}) \rangle \equiv G_{AB} |x_{ij}|^{-\Delta_A - \Delta_B} 
	\equiv \mathcal{G}_{AB},
	\label{metric}
\end{equation}
where we have defined
\begin{equation}
    x_{ij} \equiv x_{i}-x_{j}.
\end{equation}
Two-point functions of non-scalars 
scale in the same way but \( G_{AB} \) then depends on polarizations 
and the direction of \( x^{\mu}_{ij} \). The
conformal operators in eq. \ref{eq:operators} have been normalized to 
have leading \( N 
\)-independent terms in large \( N \) two-point functions (with their Hermitean 
conjugates).

The value of the two point correlator (modulo its spacetime dependence) 
of a primary operator with the Hermitean conjugate of another operator works as an inner 
product\footnote{For Hermitean operators it is just a component of the 
metric \( g_{AB} \), as seen in eq. \ref{metric}.} in the space of 
primaries. The same quantity for any operators we 
call ``overlap'', by abuse of terminology. Descendants of two
orthogonal primaries have vanishing overlap with one another. Conversely, a 
vanishing two point 
function between two operators means 
that they belong to orthogonal conformal families. Therefore, an operator
is primary if and only if it has vanishing overlap with all 
operators of lower dimension.

All operators free of derivatives are primary, simply because there are 
no operators with lower dimensions that can have non-zero overlap with 
them. But there are also numerous primaries containing derivatives, the most 
commonly known being conserved \( SO(6) \) currents and the conserved 
stress tensor. For free scalars
\begin{eqnarray}
	J_{\mu}^{IJ} & = & {1 \over N} {\mathrm{Tr}}\left\{ 
	\phi^I\partial_{\mu}\phi^J -\phi^J\partial_{\mu}\phi^I \right\} 
	= \partial^{01}\Phi^{IJ}_{\mu}-\partial^{01}\Phi^{JI}_{\mu}
	\label{current} \\
	T_{\mu\nu} & = & {{\mathrm{const}} \over N} {\mathrm{Tr}}\left\{ 
	\partial_{\mu} \phi^I \partial_{\nu} \phi^I
	- {\eta_{\mu\nu} \over 4} \partial_{\rho} \phi^I \partial^{\rho} \phi^I
	- {1 \over 2} \phi^I \partial_{\mu}\partial_{\nu} \phi^I
	+ {\eta_{\mu\nu} \over 8} \phi^I \partial^2 \phi^I\right\}
	\label{stress}
\end{eqnarray}

In our case it is also easy to construct other primaries which are 
linear combinations of terms with a single derivative.
The operators
\begin{equation}
{\mathrm{Tr}}(\phi^{I_1}\ldots\partial_\mu\phi^{I_k}\ldots\phi^{I_l}
\cdots\phi^{I_m})-
{\mathrm{Tr}}(\phi^{I_1}\ldots\phi^{I_k}\ldots\partial_\mu\phi^{I_l}
\ldots\phi^{I_m})
		\label{newops}
\end{equation}
can only have non-zero overlap with operators composed of the same 
fields. Up to permutations of the fundamental fields the only such operators of
lower dimension are
\begin{equation}
{\mathrm{Tr}}(\phi^{I_1}\ldots\phi^{I_k}\ldots\phi^{I_l}\ldots\phi^{I_m}),
        \label{primprim}
\end{equation}
which by construction have vanishing overlap with the operators 
(\ref{newops}). Therefore, expression (\ref{newops}) represents a new primary unless 
it vanishes, which it does if the derivatives happen to act on 
identical fields in cylically equivalent positions. There are also many primaries 
with more than one derivative, but 
such operators are more difficult to generate. 

The most important operators are the IR relevant and marginal 
operators, which can be 
added to the Lagrangian without destroying the UV behaviour. They 
have \( \Delta \leq 4\) and are given in table 1 in terms of their 
composition, derivative structure and \( SO(6) \) Young tableaux.  
\begin{table}[t]
\begin{center}
\begin{tabular}{|c|c|c|c|}
  \hline
  $ $ & $\Delta=2$ & $\Delta=3$ & $\Delta=4$ 
\rule[-.4cm]{0cm}{1.1cm}\\ \hline
  $\Phi^{IJ}$ & $ {\tiny \yng(2)\oplus \bullet}$ &$ $  &$ $  
\rule[-.4cm]{0cm}{1.1cm}\\ \hline 
  $\Phi^{IJK}$ & $ $ & 
${\tiny\yng(3)\oplus \yng(1)\oplus\Yvcentermath1 \yng(1,1,1)}$ &$ $ 
\rule[-.4cm]{0cm}{1.1cm}\\ \hline
  $\Phi^{IJKL}$ & $ $ & $ $ & 
$ {\tiny\yng(4)\oplus 2\; \yng(2)\oplus2\bullet\oplus2\; \Yvcentermath1\yng(2,2)\oplus2\; \yng(2,1,1)\oplus\yng(1,1)}$ 
\rule[-.4cm]{0cm}{1.1cm}\\ \hline
$(\partial^{10}\Phi^{IJ})_{\mu} $ & $ $ & ${\tiny\Yvcentermath1\yng(1,1)} $ & $ $ 
\rule[-.4cm]{0cm}{1.1cm}\\ \hline
$(\partial^{100}\Phi^{IJK})_{\mu}$ & & $ $ & ${\tiny2\; \Yvcentermath1\yng(2,1)\oplus2\; \yng(1)}$ 
\rule[-.4cm]{0cm}{1.1cm}\\ \hline
$(\partial^{\{\Sigma n=2\}}\Phi^{IJ})_{\mu\nu}$ & $ $ & $ $ & ${\tiny\yng(2)\oplus \bullet}$ 
\rule[-.4cm]{0cm}{1.1cm}\\ \hline
\end{tabular}
\end{center}
 \caption{Spectrum of relevant and marginal primaries.}
\end{table}
The table was constructed by checking the effect of the cyclic 
property of the trace, which projects out some operators and relates 
others. The primaries were then picked out. There are more marginal and
relevant gauge invariant primaries composed 
solely of scalars in the \( 
\lambda = 0 \) limit\footnote{In the canonical normalization of 
fundamental fields of eq. \ref{ScalarProp}, a marginal perturbation to a 
non-zero \( \lambda \) theory can only be achieved by adding interaction 
terms to the Lagrangian which break the original Abelian gauge symmetries 
and replace them with a deformed, non-Abelian gauge symmetry.} than in the 
supergravity limit \( \lambda 
\rightarrow \infty \). In the supergravity limit the only such scalar 
primaries are symmetric traceless tensors of \( SO(6) \) 
\cite{Witten:1998qj,Aharony:1999ti}. This indicates an intricate structure of 
branches for the moduli space of the theory, with new branches of 
conformal field theory splitting off at intermediate values of \( 
\lambda \), where some operators relevant at \( \lambda = 0 \) 
become marginal. At least we could expect that the possible IR 
limits of deformations of the theory vary strongly with the UV coupling 
\( \lambda \). Another surprise in table 1 is the last line, with 20 \( 
SO(6) \) traceless symmetric tensors, which are 
symmetric traceless in spacetime, as well as conserved. 
In AdS they are \( SO(6) \) 
charged massless spin two cousins of the graviton! If we had  
taken vector fields into account we would also have listed the vector 
contribution to the energy momentum tensor, which is an \(SO(6)\) 
scalar, and corresponds to a second AdS field with the quantum 
numbers of the graviton. At this time it is too early to say whether 
these curious facts imply that there is something seriously wrong with 
the zero coupling theory, or if they have something profound to tell 
us about stringy geometry.

Even if one has chosen a basis of mutually orthogonal primaries in the 
large \( N \) limit, there will in general be \( 1/N \) corrections to 
two-point functions which mix originally independent operators. This 
is the most basic way in which a kind of interactions appear in our 
free theory, and it is a string coupling of the order of \( 1/N \) at 
work. A few examples computed in the \( U(N) \) theory illustrates 
how the general computation consists of a combinatorial part and an 
analytic part, which takes care of the polarization dependence of the 
two-point function. 
\begin{eqnarray}
\left\langle\Phi^{123}(x)\Phi^{123}(0)\right\rangle &=&\frac{1}{N^3}
<:\left[{\mathrm{Tr}}(\phi^{1}\phi^{2}\phi^{3})\right](x):
:\left[{\mathrm{Tr}}(\phi^{1}\phi^{2}\phi^{3})\right](0):>\nonumber\\
&=&\frac{1}{N^2}|x|^{-6}
\end{eqnarray}
\begin{eqnarray}
\lefteqn{\left\langle\left[\Phi^{12}\Phi^{13}\right]\!(x)\:\Phi^{1231}(0)\right\rangle}\nonumber\\
&=&\frac{1}{N^4}\left\langle :\left[{\mathrm{Tr}}(\phi^{1}\phi^{2}){\mathrm{Tr}}(\phi^{1}\phi^{3})\right](x):
:\left[{\mathrm{Tr}}(\phi^{1}\phi^{2}\phi^{1}\phi^{3})\right](0):\right\rangle\nonumber\\
&=&\frac{1}{N}|x|^{-8}+\frac{1}{N^3}|x|^{-8}
\end{eqnarray}
\begin{eqnarray}
\lefteqn{\left\langle J^{12}_\mu(x)J^{12}_\nu(y)\right\rangle}\nonumber\\
&=&\frac{1}{N^2}\left\langle :\left[{\mathrm{Tr}}(\phi^{1}\partial_\mu\phi^{2}
-\phi^{2}\partial_\mu\phi^{1})\right](x):
:\left[{\mathrm{Tr}}(\phi^{1}\partial_\nu\phi^{2}
-\phi^{2}\partial_\nu\phi^{1})\right](y):\right\rangle\nonumber\\
&=&2|x-y|^{-2}\frac{\partial}{\partial x^\mu}
\frac{\partial}{\partial y^\nu}|x-y|^{-2}
-2\frac{\partial}{\partial y^\nu}|x-y|^{-2}
\frac{\partial}{\partial x^\mu}|x-y|^{-2}
\end{eqnarray}
The combinatorial calculation involves counting 
how many closed index lines are formed between the two operators in the double line 
representation, to give the appropriate \( N \)-dependence. Note that 
the convention of normal ordering operators just means that 
propagators should not return to the same operator. If there 
are several ways of saturating the operators with propagators, they 
should be added, and will in general give rise to a polynomial 
dependence on \( 1/N \).  

To give the two-point function a simple and physical form, one should diagonalize 
the mixing matrix. Because primary operators can only mix with operators 
of the same 
dimension, and two operators that mix also have to consist of equal numbers 
of all fundamental fields, the mixing problem can be reduced to a 
block diagonal form. Only finite-dimensional 
diagonalizations are needed to find the exact propagator.

Single-trace operators may also mix with multiple-trace operators, i.e. 
products of single-trace operators. A natural AdS interpretation of 
such operators is as multi-string states, but it is somewhat puzzling 
that such products of independent operators with equal argument should 
play a special role, in addition to being limits of products of 
unequal arguments. Presumably the normal ordering needed to 
regularize the product can be interpreted in AdS as a way of 
binding the two strings to each other in the radial direction 
(which in the AdS correspondence is related to the boundary theory 
scale \cite{Susskind:1998dq,Peet:1999wn}). 

If one includes the multiple-trace operators among the operators that 
can mix, one gets larger matrices to diagonalize, but still of finite dimension, 
by the same argument as before. The resultant 
diagonalized full 
propagator propagates \( N \)-dependent linear combinations of 
single-trace and multi-trace operators, without mixing among these superpositions.
Their 
dimensions are all unchanged, and \( N \)-independent. This result 
about \( N \)-independence at \( \lambda = 0 \) 
sharpens the assertion in \cite{D'Hoker:1999jp} about the behaviour of the 
dimensions of multi-trace operators at weak coupling. In contrast, 
the \emph{strong} 't Hooft coupling result of D'Hoker et al \cite{D'Hoker:1999jp}
is that the dimensions of multi-trace operators do shift.

The block overlap matrices should become degenerate for some finite \( N 
\), depending on the block. This is because there are linear dependencies 
among the naive states \cite{Sundborg:1999ue}, known
as a string exclusion principle 
\cite{Maldacena:1998bw}. The determinants of the block 
overlap matrices are polynomials 
in \( 1/N \), so the smallest root of each determinant sets the value of 
\( N \) for which \( 1/N \) perturbation theory breaks down in the given 
block. 

\section{Operator products and string vertices}
\label{Opsv}

Essentially all string theory interactions may be derived from 
three-string vertices, roughly because all string diagrams can be 
constructed by sewing together pant diagrams (which carry the 
three-string structure). In many approaches additional contact terms 
are also needed, but their role is mainly to make sense of analytic 
continuations. Similarly, in conformal field theory we expect the 
three-point functions (and analytic continuation) to be enough to 
calculate any correlation function. The three-point functions contain 
essentially the same information as the operator product expansion, 
which completely characterizes 
the theory if conformal bootstrap \cite{Pol} works as in two 
dimensions \cite{Belavin:1984vu}. For general operators the 
three-point function
\begin{eqnarray}
\lefteqn{ \left\langle A(x_{1}) B(x_{2}) C(x_{3})\right\rangle}\nonumber\\
&&=\frac{C_{ABC}}{|x_{12}|^{\Delta_A+\Delta_B-\Delta_C}
|x_{31}|^{\Delta_A+\Delta_C-\Delta_B}
|x_{23}|^{\Delta_C+\Delta_B-\Delta_A}}
\equiv\mathcal{C}_{ABC},
\end{eqnarray}
where spacetime dependence is included in \(\mathcal{C}_{ABC} \), which 
as \( C_{ABC}\)
typically depends on the spins of the operators and the relative orientations 
of the \(x_{ij}^{\mu}\). The general operator product expansion
\begin{equation}
A(x) B(y)\sim \sum_{D}
C_{AB}^{\;\;\;\;\;D} D(y)|x-y|^{\Delta_D-\Delta_A -\Delta_B}= 
\sum_{{\mathcal{D}}}{\mathcal{C}}_{AB}^{\;\;\;\;\;D} D(y)
\label{eq:OPE}
\end{equation}
is formally related to the three-point function through
\begin{equation}
    \mathcal{C}_{AB}^{\;\;\;\;\;D}\equiv 
    \mathcal{C}_{ABC}\:\mathcal{G}^{CD},
    \label{eq:raising}
\end{equation}
with \( \mathcal{G}^{AB} \) the inverse of the propagator
\( \mathcal{G}_{AB} \).

The \( n \)-point functions are constrained by the requirement that all 
fundamental fields should be joined by a propagator to a fundamental 
field in another operator (see fig. 1). This implies that all non-zero
correlation functions contain an even number of fundamental fields.
\begin{figure}[t]
\begin{center}
\leavevmode
\vbox{
\epsfxsize=5cm
\epsffile{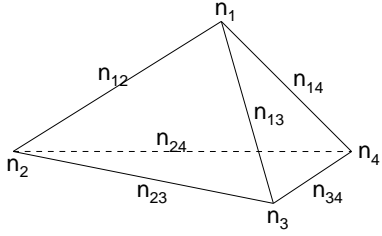}}
\end{center}
\caption{A schematic description of a four-point function as a 
tetrahedron. To resolve the \( N \)-dependence the order of the fields 
in each operator also has to be taken into account.}
\label{pyramid}
\end{figure}
Furthermore, any \( n \)-point function can be 
represented by an \( n \)-hedron for each kind of fundamental field (see 
fig. \ref{pyramid}). There 
are \( n^{I}_{i} \) fields \( \phi^{I} \) at the \( i \)-th corner, 
and \( n^{I}_{ij} \) propagators of \( \phi^{I} \) along the  \( ij \equiv 
ji \) 
edge. We must have \(  n^{I}_{i} = n^{I}_{i1} +  \ldots +  n^{I}_{in} 
\) and
\begin{equation}
	n^{I}_{ij}= {1 \over n-2} (n^{I}_{i}+n^{I}_{j} - {{n^{I}_{1} +  
	\ldots +  n^{I}_{n}} \over n-1})
	\label{eq:edge}
\end{equation}
For the three-point function, \( n=3 \), a non-negative number of 
propagators \( n^{I}_{ij}\geq 0 \) implies triangle inequalities 
\begin{equation}
	n^{I}_{\pi(1)} \leq n^{I}_{\pi(2)} + n^{I}_{\pi(3)}
	\label{eq:triangleineq}
\end{equation}
for any permutation \( \pi \).

The underlying reason for the rules above is that we are dealing with 
a free theory, which is invariant to independent shifts of all the 
fundamental scalars. The corresponding conserved currents are \( 
J_{\mu\beta}^{I\alpha} = \partial_{\mu}\phi^{I \alpha}_{\beta} \), which are 
not gauge singlets, and thus not among the operators we would otherwise 
consider.

In our case we have a free theory, and the OPE can be obtained by 
first applying Wick's theorem and then Taylor expanding the result. For 
example
we have
\begin{eqnarray}
\lefteqn{	{1 \over N} :\!\!{\mathrm{Tr}}\{\phi^2 (x_{i})\}\!\!: \;{1 \over N} 
	:\!\!{\mathrm{Tr}}\{\phi^2 (x_{j})\}\!\!: \;= } \nonumber \\
	&& = {1 \over N^2} :{\mathrm{Tr}}\{\phi^2 (x_{i})\}
	{\mathrm{Tr}}\{\phi^2 (x_{j})\}:
	+{4 \over N^2 |x_{ij}|^2}:{\mathrm{Tr}}\{\phi(x_{i}) \phi(x_{j})\}:
	+{4 \over |x_{ij}|^4} \nonumber\\
	&&={1 \over N^2} \sum_{n}{(x_{ij})^{\mu_{1}}\ldots(x_{ij})^{\mu_{n}} \over n!}
	:{\mathrm{Tr}}\{\partial_{\mu_{1}}\ldots\partial_{\mu_{n}}
	\phi^2 (x_{j})\}{\mathrm{Tr}}\{\phi^2 (x_{j})\}:\nonumber\\
        &&+{4 \over N^2 |x_{ij}|^2}\sum_{n}{(x_{ij})^{\mu_{1}}\ldots(x_{ij})^{\mu_{n}} \over n!}
	:{\mathrm{Tr}}\{\partial_{\mu_{1}}\ldots\partial_{\mu_{n}}\phi(x_{j}) 
	\phi(x_{j})\}:
	+{4 \over |x_{ij}|^4}.
	\label{eq:WickOPE}
\end{eqnarray}
Terms proportional to the unit operator do not contribute to 
three-point functions, but as we will see explicitly in section 
\ref{Cd}, they are 
essential for the \( 1/N 
\)-expansion to produce disconnected diagrams. Such diagrams are of course 
needed if the expansion is to be interpreted as a perturbative expansion of 
string theory.

Since the model is essentially a trivial free field theory, only studied from the 
special perspective of its gauge-invariant local operators, we might 
worry that the corresponding string theory is also trivial. In 
particular, we might ask if there are only lowest order, 
\( 1/N \), string interactions. Could it be that diagonalization of the full
two-point function is enough, and absorbs all other \( N \)-dependence? 
For flat space amplitudes, such 
behaviour would be impossible in an interacting theory because of 
\( S \)-matrix unitarity\footnote{It
generates an order \( g^2 \) imaginary part from an interaction of 
order \( g \), etc\ldots}. In the present theory, we do not have an ordinary \( 
S \)-matrix, neither in the four-dimensional Minkowski space because of 
conformal invariance, nor in the five-dimensional
gravitational picture because of the AdS background, 
so this argument does not necessarily 
apply. To resolve the issue we have found a three-point function 
with only higher order interactions, and checked that diagonalization 
of the two-point functions of the three operators cannot reduce the 
interaction to order \( 1/N \), the coupling strength of the 
fundamental interactions. 
This demonstrates that the theory is a highly non-trivial interacting 
theory even at zero 't Hooft coupling (i.e. for tensionless strings). 

\begin{figure}[t]
\begin{center}
\leavevmode
\vbox{
\epsfxsize=12cm
\epsffile{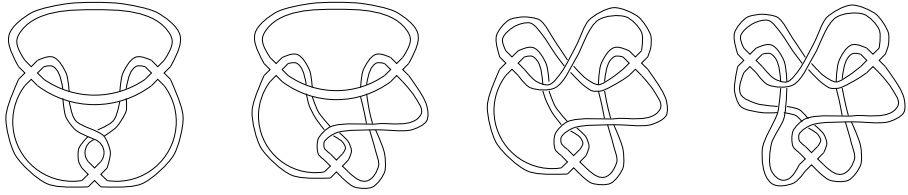}}
\end{center}
\caption{Leading diagrams contributing to the correlator (\ref{eq:non-trivial}).}
\label{3pfcns}
\end{figure}

Consider the correlation function
\begin{equation}
    \langle \Phi^{1212}(x_{1})\Phi^{2323}(x_{2})\Phi^{3131}(x_{3}) \rangle
    \label{eq:non-trivial}
\end{equation}
among single-trace operators. The leading contributions are shown in 
fig. \ref{3pfcns}, and they are of order \( 1/N^3 \). The 
three operators involve different fields and cannot mix pairwise with each other. 
Thus, no diagonalization of single-trace operators can give this three-point 
function from a \( 
1/N \) vertex and \( 1/N \)-corrected external states.

By diagonalization among the full set of gauge invariant operators, 
including multi-trace operators, it is possible to get terms like the leading 
contribution as a result of an admixture of double-trace operator in 
the external state, 
but again vertices of higher order than \( 1/N \) are needed to couple 
to the remaining two states. 
To see this factorize a diagram of fig. \ref{3pfcns} into a first factor 
consisting of  \( 1/N^2 \) three-point vertices \( \langle 
\Phi^{IJIJ}\Phi^{JKJK}[\partial^{\{n\}}\Phi^{KI}\partial^{\{m\}}\Phi^{KI}]\rangle \) 
between two external single-trace operators and two-trace operators 
and a second factor consisting of a \( 1/N \) mixing 
\( \langle [\partial^{\{n\}}\Phi^{KI}\partial^{\{m\}}\Phi^{KI}] 
\Phi^{KIKI}\rangle \) 
of such two-trace 
operators with the remaining external single-trace operator. This 
example indicates that it may be possible to write the full theory 
in terms of completely diagonalized local operators, but also that 
three-point vertices of higher order in \( 1/N \) are needed.

\section{CFT crossing symmetry and string amplitude duality}
\label{Cd}

\begin{figure}[t]
\begin{center}
\leavevmode
\vbox{
\epsfxsize=6cm
\epsffile{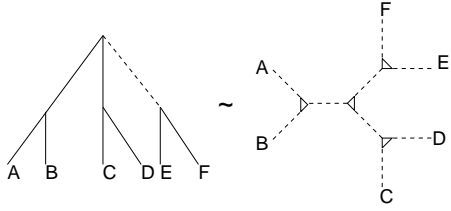}}
\end{center}
\caption{Diagrams showing two possible expansions of a six-point 
functoion given in eq. \ref{eq:six-point}.}
\label{fig3}
\end{figure}

The OPE, eq. \ref{eq:OPE}, can be used inside correlation 
functions in several ways depending on which distances are assumed to be small, 
and at what points the operator products are to be inserted. By a 
sequence of expansions an \( n \)-point function can be reduced to 
operator product coefficients \( {\mathcal{C}}_{AB}^{\;\;\;\;\;D} \)
joined by operator two-point 
functions \( \left\langle C(x_i) D(x_j) \right\rangle \), all multiplied 
together and summed over all possible propagating operators. 
More symmetrically, the \( n \)-point function may be expressed in terms 
of two-point functions and the amputated three-point function \( 
\mathcal{C}^{ABC}\), obtained by multiplying the three-point function 
by inverse two-point functions. For example a six-point function can 
be written as 
\begin{eqnarray}
    \lefteqn{
    {\mathcal{C}}_{AB}^{\;\;\;\;\;B'}
    {\mathcal{C}}_{DC}^{\;\;\;\;\;C'}
    {\mathcal{C}}_{FE}^{\;\;\;\;\;E'}
    {\mathcal{C}}_{B'C'}^{\;\;\;\;\;C''}{\mathcal{G}}_{C''E'}}
    \nonumber \\
   &=&
    {\mathcal{G}}_{{AA}_{1}}{\mathcal{G}}_{{BB}_{1}}
    {\mathcal{G}}_{{CC}_{1}}{\mathcal{G}}_{{DD}_{1}}
    {\mathcal{G}}_{{EE}_{1}}{\mathcal{G}}_{{FF}_{1}}
    {\mathcal{C}}^{A_{1}B_{1}B'}
    {\mathcal{C}}^{C_{1}D_{1}C'}
    {\mathcal{C}}^{E_{1}F_{1}E'}
    {\mathcal{G}}_{{B'}{B}_{1}'}
    {\mathcal{G}}_{{C'}{C}_{1}'}
    {\mathcal{G}}_{{E'}{E}_{1}'}
    {\mathcal{C}}^{B_{1}'C_{1}'E_{1}'}
    \label{eq:six-point}
\end{eqnarray}
Diagrammatically this may be drawn as in fig. \ref{fig3}. In 
the second way of writing the six-point function additional internal spacetime 
points serving as arguments of two-point functions and amputated 
three-point functions are introduced. If only the expansions converge 
the locations of these points are arbitrary.

No
single sequence of expansions converges for all positions of 
the operators, but one can hope that different sequences should converge 
in complementary regions in the space of operator positions, and yield 
continuations of each other. Interpreted in terms of an AdS string \( S 
\)-matrix this would mean that the amplitude could be expanded in 
kinematic invariants in many different ways that are continuations of 
one another. But this is just the kind of scattering ``duality" that was 
the origin of string theory \cite{Veneziano:1968yb}, and which is intuitively reasonable 
when the amplitude is viewed as a result of a path integral over string world 
sheets with no interaction points (in contrast to corresponding 
amplitudes for point-particle theory). In our case we do not have an 
actual world-sheet picture, just sets of free particle propagators that 
can span a surface thanks to large \( N \) counting. But by examining a 
four-point function we can see explicitly how  string scattering duality 
emerges from the OPE of conformal field theory.

If we insert the OPE eq. \ref{eq:WickOPE} twice into the four-point function 
of normalized quadratic traces of \( U(N) \) scalars
\begin{equation}
{1 \over N^4}\left\langle : \!\!{\mathrm{Tr}}\:\phi^2 (x_{1})\!\! : \; 
: \!\!{\mathrm{Tr}}\:\phi^2 (x_{2})\!\! :
	\; : \!\!{\mathrm{Tr}}\:\phi^2 
    	(x_{3})\!\! : \;
	: \!\!{\mathrm{Tr}} \:\phi^2 (x_{4})\!\! : \right\rangle 
	\label{eq:GFcn}    
\end{equation}
we get
\begin{eqnarray}
    \lefteqn{{1 \over N^4}\left\langle : \!\!{\mathrm{Tr}}\:\phi^2 (x_{1})\!\! : \;
     : \!\!{\mathrm{Tr}}\:\phi^2 (x_{2})\!\! :
	\; : \!\!{\mathrm{Tr}}\:\phi^2 (x_{3})\!\! : \;
    	 : \!\!{\mathrm{Tr}} \:\phi^2 (x_{4})\!\! : \right\rangle} 
	 \\
	&=& {16 \over |x_{12}|^4 |x_{34}|^4} +{1 \over N^4} 
	\sum_{n,m}{(x_{12})^{\mu_{1}}\ldots(x_{12})^{\mu_{n}} \over n!}
	{(x_{34})^{\nu_{1}}\ldots(x_{34})^{\nu_{m}} \over m!} \times \nonumber \\
	&&\times \biggl(\left\langle\; : \!\!\bigl[{\mathrm{Tr}}\;\partial_{\mu_{1}}\ldots\partial_{\mu_{n}}
	        \phi^2 \;{\mathrm{Tr}}\:\phi^2 \bigr] (x_{2})\!\! : \;
	    : \!\!\left[{\mathrm{Tr}}\; \partial_{\nu_{1}}\ldots\partial_{\nu_{m}}
	                \phi^2 \;{\mathrm{Tr}} \:\phi^2\right]  (x_{4})\!\! : \;\right\rangle 
		\nonumber \\ 
	&&\;\;\;\;\;+\;{16 \over |x_{12}|^2 |x_{34}|^2}
	\left\langle\; : \!\!\left[{\mathrm{Tr}}\: \phi \,\partial_{\mu_{1}}\ldots\partial_{\mu_{n}}
	        \phi  \right] (x_{2})\!\! : \;
	    : \!\! \left[{\mathrm{Tr}}\: \phi \,\partial_{\nu_{1}}\ldots\partial_{\nu_{m}}
	        \phi  \right] (x_{4})\!\! : \;\right\rangle \biggr) ,
	\nonumber		\label{doubleOPE}
\end{eqnarray}
for small \(x_{12}\) and \(x_{34}\) relative to \(x_{23}\) and \(x_{14}\). 
The first term comes from the terms proportional to the unit operator 
in the OPEs, corresponds to disconnected diagrams. The second line on 
the right-hand side corresponds to the propagation of quartic operators, 
and consists of one connected and two disconnected pieces 
(corresponding to propagation of double-trace operators). Finally, the 
last line consists of connected diagrams propagating quadratic 
operators. A direct Taylor 
expansion of the Green function (\ref{eq:GFcn}) gives the same result as this double 
OPE, and the regions of convergence are the same. 
There are three different ways of combining the four 
external operators into two pairs, each yielding a different expansion of the 
same Green function. Therefore, the full Green function
can be obtained as a continuation of expansions in any such channel. 
This is string scattering duality for the 
corresponding AdS amplitude.

The basic reason for the above duality appears to be that products of 
normal ordered operators are associative. Presumably the 
associativity can be 
used to prove rigorously many of the formal identities discussed above relating 
\(n\)-point functions, OPE coefficients, three-point functions and 
two-point functions.

\section{Discussion}
\label{Disc}

We have used the AdS/CFT conjecture as a tool to tentatively define string 
theory in \( AdS_{5} \times S^5 \) with a Ramond-Ramond background. 
Although we have used the correspondence outside the region where it has 
been tested, at small 't Hooft coupling, we have found that such a
definition gives rise to a non-trivial interacting theory with the 
fundamental
properties of a string theory, like duality of scattering amplitudes. We 
have tested a simple four-point amplitude and verified that CFT crossing 
symmetry gives rise to the desired behaviour. We have listed marginal and 
relevant primary operators composed of scalars and found that there are 
more such operators at small 't Hooft coupling than at large, indicating 
a complicated phase diagram of IR deformations of \( {\cal N}=4 \) SYM. 
In string theory we expect a large number of backgrounds which are 
asymptotically AdS. 

Surprisingly we have found several marginal traceless symmetric tensors, 
which correspond to massless spin 2 particles in AdS. Somehow, the 
extremely stringy tensionless limit involves several geometries 
interacting with each other. It remains to be seen if this is a defect 
which can only be cured by a perturbation to non-zero tension, or if it 
is a consistent and perhaps even a characteristic property of string 
theory at extremely short distances.

Furthermore we have argued that the theory in the limit of vanishing 't 
Hooft coupling allows a complete diagonalization of the string 
propagator. Nevertheless, we have found the theory to be a complicated 
interacting theory with interactions of all orders in \( 1/N \). 

A puzzling question is if the purely combinatorial \( 1/N \) expansion in 
the zero coupling theory, which by large \( N \) lore is 
the genus expansion of string theory, can be related to sums of 
intermediate single- and multiple-string states. In particular, one would 
expect string loops to correspond to integrals or sums over {\it all\/} 
multi-string intermediate states (composed of arbitrarily many 
fundamental fields) that can couple to the external states. For 
loop sums to equal the 
combinatorial sums there apparently have to be enormous cancellations, 
since the number of fundamental propagators in the sums is 
bounded by expressions 
like eq. \ref{eq:edge}. Perhaps such cancellations are typical of 
extremely holographic systems.

\bigskip

\begin{flushleft}
We would like to thank H.\ Hansson for discussions and B. Brinne for 
helping us to find the tools to draw Young tableaux. The work 
of B. S. was financed by 
the Swedish Science Research Council.
\end{flushleft}


\begin{thebibliography}{99}
 

\bibitem{'tHooft:1974jz}
G.~'t Hooft,
Nucl.\ Phys.\  {\bf B72} (1974) 461.

\bibitem{Maldacena:1997re}
J.~Maldacena,
Adv. Theor. Math. Phys. {\bf 2} (1998) 231
hep-th/9711200.

\bibitem{Gubser:1998bc}
S.S.~Gubser, I.R.~Klebanov and A.M.~Polyakov,
Phys. Lett. {\bf B428} (1998) 105
hep-th/9802109.

\bibitem{Witten:1998qj}
E.~Witten,
Adv. Theor. Math. Phys. {\bf 2} (1998) 253
hep-th/9802150.

\bibitem{Aharony:1999ti}
O.~Aharony, S.S.~Gubser, J.~Maldacena, H.~Ooguri and Y.~Oz,
hep-th/9905111.

\bibitem{Kallosh:1998qs}
R.~Kallosh and A.~Rajaraman,
Phys.\ Rev.\  {\bf D58} (1998) 125003
[hep-th/9805041].

\bibitem{null}
F.~Lizzi, B.~Rai, G.~Sparano and A.~Srivastava,
Phys.\ Lett.\  {\bf B182} (1986) 326;

A.A.~Zheltukhin, Sov.\ J.\ Nucl.\
Phys.\ {\bf 48} (1988) 375;


J.~Barcelos-Neto and M.~Ruiz-Altaba,
Phys.\ Lett.\  {\bf B228} (1989) 193;


J.~Gamboa, C.~Ramirez and M.~Ruiz-Altaba,
Nucl.\ Phys.\  {\bf B338} (1990) 143;



J.~Isberg, U.~Lindstrom, B.~Sundborg and 
G.~Theodoridis,
Nucl.\ Phys.\  {\bf B411} (1994) 122
[hep-th/9307108];


H.~Gustafsson, U.~Lindstrom, P.~Saltsidis, 
B.~Sundborg and R.~van Unge,
Nucl.\ Phys.\  {\bf B440} (1995) 495
[hep-th/9410143].



\bibitem{Pesando:1998fv}
I.~Pesando,
JHEP {\bf 9811} (1998) 002
[hep-th/9808020].

\bibitem{Kallosh:1998nx}
R.~Kallosh and J.~Rahmfeld,
Phys.\ Lett.\  {\bf B443} (1998) 143
[hep-th/9808038].

\bibitem{Kallosh:1998ji}
R.~Kallosh and A.~A.~Tseytlin,
JHEP {\bf 9810} (1998) 016
[hep-th/9808088].

\bibitem{Berkovits:2000fe}
N.~Berkovits,
hep-th/0001035.

\bibitem{Balasubramanian:1999ri}
V.~Balasubramanian, S.~B.~Giddings and A.~Lawrence,
JHEP {\bf 9903} (1999) 001
[hep-th/9902052].

\bibitem{Giddings:1999qu}
S.~B.~Giddings,
Phys.\ Rev.\ Lett.\  {\bf 83} (1999) 2707
[hep-th/9903048].

\bibitem{Thorn:1991fv}
C.~B.~Thorn,
hep-th/9405069.

\bibitem{Sundborg:1999ue}
B.~Sundborg,
hep-th/9908001.

\bibitem{Susskind:1998dq}
L.~Susskind and E.~Witten,
hep-th/9805114.

\bibitem{Peet:1999wn}
A.~W.~Peet and J.~Polchinski,
Phys.\ Rev.\  {\bf D59} (1999) 065011
[hep-th/9809022].

\bibitem{D'Hoker:1999jp}
E.~D'Hoker, S.~D.~Mathur, A.~Matusis and L.~Rastelli,
hep-th/9911222.

\bibitem{Maldacena:1998bw}
J.~Maldacena and A.~Strominger,
JHEP {\bf 9812} (1998) 005
[hep-th/9804085].

\bibitem{Pol}
A.~M.~Polyakov,
Sov.\ Phys.\ JETP {\bf 39} (1974) 10.

\bibitem{Belavin:1984vu}
A.~A.~Belavin, A.~M.~Polyakov and A.~B.~Zamolodchikov,
Nucl.\ Phys.\  {\bf B241} (1984) 333.

\bibitem{Veneziano:1968yb}
G.~Veneziano,
Nuovo Cim.\  {\bf A57} (1968) 190.



\end{thebibliography}
\end{document}